\documentclass[journal=accacs,manuscript=article]{achemso}

\usepackage[version=4]{mhchem} % Formula subscripts using \ce{}
\usepackage{color}
\usepackage{ulem}

\usepackage{amsmath}
\usepackage{pdfpages}
\author{Lance Kavalsky}
\affiliation[Carnegie Mellon University]
{Department of Mechanical Engineering, Carnegie Mellon University, Pittsburgh, Pennsylvania 15213, USA}

\author{Venkatasubramanian Viswanathan}
\affiliation[Carnegie Mellon University]
{Department of Mechanical Engineering, Carnegie Mellon University, Pittsburgh, Pennsylvania 15213, USA}
\email{venkvis@cmu.edu}
\title[An \textsf{achemso} demo]
  {Robust Active Site Design of Single Atom Catalysts for Electrochemical Ammonia Synthesis}
  

\begin{document}

\begin{tocentry}

% Some journals require a graphical entry for the Table of Contents.
% This should be laid out ``print ready'' so that the sizing of the

% Inside the \texttt{tocentry} environment, the font used is Helvetica
% 8\,pt, as required by \emph{Journal of the American Chemical

% The surrounding frame is 9\,cm by 3.5\,cm, which is the maximum
% permitted for  \emph{Journal of the American Chemical Society}
% graphical table of content entries. The box will not resize if the
% content is too big: instead it will overflow the edge of the box.

% This box and the associated title will always be printed on a
% separate page at the end of the document.                 

% \begin{figure}
  \includegraphics[width=9cm]{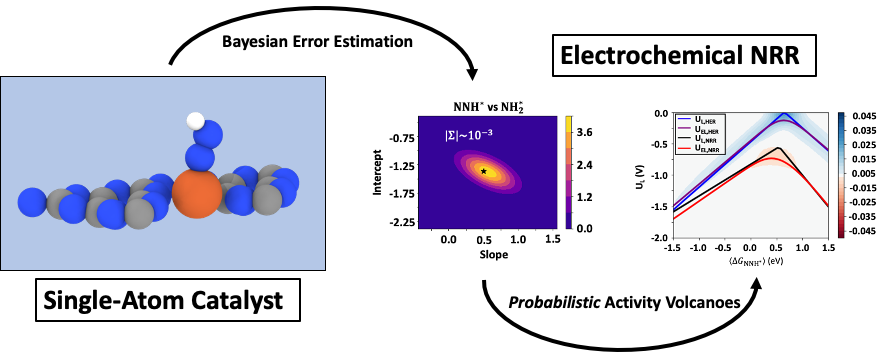}
% \end{figure}

\end{tocentry}

\begin{abstract}
Electrochemical ammonia synthesis forms a key part of sustainable chemicals synthesis.  Single-atom catalysts have emerged as a promising class of electrocatalysts that could be capable of electrochemically reducing nitrogen into ammonia.  The analysis of electrochemical reduction of nitrogen is complicated by multiple mechanistic pathways and the competing hydrogen evolution reaction. The identified pathways using thermodynamic analysis based on DFT calculations is strongly dependent on the choice of the exchange correlation functional. In this work, we provide a computational methodological framework using the single-atom systems as an example material class for ammonia synthesis that is robust towards parameter selection. Applying this to Pt$_1$/g-\ce{C3N4}, Ru$_1$/g-\ce{C3N4}, and Fe$_1$/g-\ce{C3N4}, we generate ensembles of limiting potentials, using the ensemble of functionals collected via Bayesian Error Estimation Functionals (BEEF), to robustly predict catalytic activity. We then extend this to study the scaling between NRR reaction intermediates and use it to identify that \ce{NNH}* as the best descriptor for these relations. In addition, a procedure to investigate selectivity is outlined, and a more robust way to analyze the selectivity-activity trade-off is presented. For this single-atom material class, we find choosing catalysts that lie on the strong binding leg of the activity volcano are worth further exploration.  Given the ease of integration of the proposed method with minimal additional computational cost, we believe this should become a routine part of analysis workflow for multi-electron electrochemical reactions.
% three main things from the paper

\end{abstract}

\section{Introduction}
% NRR important reaction.  Electrochemical one possible route.
% Competing reaction, HER, many reaction intermediates.

% Talk about UQ of electrochemical reactions and how that affects mechanistic conclusions.

% This work aims to build the foundational understanding.
% We find ... (3 main conclusions of the paper)
                
In the path towards sustainable chemical synthesis, ammonia (\ce{NH3}) production is an important challenge due to its significant \ce{CO2} emissions of over 300 million metric tonnes  \cite{Guo2019RecentReduction}. This is a consequence of the widespread use of the Haber-Bosch process for synthesis, which demands harsh conditions in order to thermally drive the kinetics of the reaction \cite{Haber1905UberElementen.}. Electrochemical synthesis emerges as a promising alternative \cite{Cui2018AConditions}. In this case, \ce{N2} is reduced to \ce{NH3} in an electrochemical cell where the thermal driving force from the extreme conditions of the Haber-Bosch process is substituted for an applied electric potential which can be generated through a sustainable sources such as solar or wind \cite{Liu2020CurrentConditions}. %\vv{cite a review}

However, the electrochemical nitrogen reduction reaction (NRR) has numerous challenges \cite{Suryanto2019ChallengesAmmonia,Foster2018CatalystsAmmonia}. Firstly, poor activities have been predicted for NRR due to scaling relationships \cite{Skulason2012AReduction,Montoya2015TheRelations}. These scaling relations impose limits on the maximum achievable limiting potential by reducing the degrees of freedom in optimizing the free energy landscape. This is further complicated as the reaction is multi-step in nature, and multiple possible mechanisms have been put forward \cite{Suryanto2019ChallengesAmmonia}. Being able to confidently report the mechanism on a given surface is crucial in not only predicting NRR activity, but also in the development of design principles to guide future studies. At this point, consensus over the dominant mechanisms remains unclear\cite{Suryanto2019ChallengesAmmonia}. The underlying mechanism can have significant influence on predicted limiting potentials and scaling relations which can then dramatically alter conclusions drawn about a given catalyst. Another important consideration is competition with the hydrogen evolution reaction (HER) yielding poor faradaic efficiencies \cite{Singh2019StrategiesSynthesis}. In order to maximize yield, it is important to carefully select a catalyst material that can promote \ce{NH3} production while minimizing the competing \ce{H2} production. 
% Historically, Fe and Ru have been recognized as two of the best catalysts towards NRR, but still suffer from poor selectivity \textbf{[ref]}. 

Single-atom catalysts are an exciting new class of materials due to their high metal utilization and tunable catalytic properties \cite{DeRita2019StructuralReactivity,Ren2019UnravelingCatalyst,Li2019SupportedCatalysis,Exner2020BeyondSelectivity}. In this context of ammonia synthesis, it was previously reported that careful design of the active site via substitution in model bulk systems could improve thermochemical NRR performance \cite{Singh2018ComputationalSynthesis}. Translating this to electrochemical NRR, single Ru atoms deposited on N-doped carbon exhibited a very high yield of 120.9 $ \mu \textrm{g}_{\ce{NH3}} \textrm{mg}_{cat}^{-1} \textrm{h}^{-1}$ \cite{Geng2018AchievingCatalysts}, and Fe$_{SA}$-N-C was reported to have among the highest faradaic efficiency to date of 56.55\% \cite{Wang2019OverPotential}. Continuing on this trajectory of carbon and nitrogen based supports, graphitic carbon nitride (g-\ce{C3N4}) has drawn attention due to its unique geometric structure leading to naturally forming cavities, and some recent efforts have focused on scalable synthesis of this material \cite{Lakhi:2017aa,Yadav2020FacileNanolayers,Villalobos2020Large-scaleFilms}. Moreover, many single-atom species have already been synthesized on this support including Fe, Ru, and Pt \cite{Chen2017StabilizationNitride,An2018High-DensityProcesses,Tian2018Temperature-ControlledCatalyst}. This has inspired previous computational investigations into the performance of single-atoms on the surface, but so far they have come to differing conclusions in terms of relative NRR catalytic performance \cite{Yin2019Pt-embeddedSynthesis,Liu2019BuildingCatalysts,Chen2019ComputationalElectroreduction}. Thus, a robust analysis framework with uncertainty estimates is crucial to handle this challenging reaction.

% BEEF -- uncertainty quantification. applied to robustly draw conclusions depending on functional.  demonstrated for simple reactions.  
The significant influence of the choice of exchange-correlation functional on the DFT predictions for NRR is recognized, however a methodology for addressing this quantitatively is lacking \cite{Kepp2018AccuracySynthesis}. One approach to quantify the uncertainty arising from the choice of exchange-correlation functional (XC) is through the use of the BEEF-vdW functional \cite{Wellendorff2012DensityEstimation}. One of the first studies incorporating this approach investigated the catalytic activity towards ammonia synthesis of a collection of bulk surfaces for a dissociative mechanism \cite{Medford2014AssessingRates}. Since then, this uncertainty quantification method has been successfully implemented to a variety of applications including chlorine evolution \cite{Sumaria2018QuantifyingEvolution} and electrode-electrolyte interfacial behaviour for both oxygen reduction \cite{Vinogradova2018QuantifyingInterfaces} and CO reduction \cite{Bagger2019ElectrochemicalInterface}. Thus, this approach has demonstrated a promising method to robustly draw conclusions about a given catalyst candidate.

In this work, we propose a methodological framework for investigating the NRR catalytic ability of single-atom systems with the inclusion of uncertainty estimation and propagation.  We apply this framework using an example class of materials -  Pt$_1$/g-\ce{C3N4} (Pt$_1$), Ru$_1$/g-\ce{C3N4} (Ru$_1$), and Fe$_1$/g-\ce{C3N4} (Fe$_1$) to robustly quantify the activity and selectivity accounting for the uncertainty associated with density functional theory (DFT) calculations. Through this, we are able to generate an ensemble of limiting potentials for each of the three systems to provide more realistic predictions in terms of activity. We then study the inherent scaling relationships of the systems towards NRR, and find that from incorporating uncertainty, the NNH* intermediate is the best descriptor of these relations. Finally, we investigate the issue of selectivity, and show that due to the scaling of NNH* with H*, a selectivity-activity trade-off emerges. Based upon this analysis, the design criteria emerges that for this material class is that catalysts which lie on the strong binding leg of the volcano are worth further exploration. In summary, this work aims to simultaneously provide a deeper understanding of these systems while laying out the foundations for future single-atom system investigations for ammonia synthesis.

\section{Methodology}

\subsection{Computational Parameters}
% General DFT
Spin-polarized DFT calculations were conducted using the GPAW package \cite{ISI:000226735900040} through the Atomic Simulation Environment package \cite{ase-paper}. Ion-electron interactions were treated using the Projector Augmented Wave approach \cite{PhysRevB.50.17953}. For all calculations a grid spacing of 0.16 \AA{} and a 4 $\times$ 4 $\times$ 1 Monkhorst-Pack k-mesh were used \cite{Monkhorst1976SpecialIntegrations}. Geometric relaxations of each of the structures were done until a force criterion of $<$0.05 ev/\AA{} was met. Since the basal plane of g-\ce{C3N4} is hydrophobic \cite{Xu2017TheAmphiphile}, solvation effects were assumed to be negligible. To avoid interactions among images in the z-direction, a vacuum spacing of 20 \AA{} was introduced. To improve self-consistent field convergence, Fermi smearing was applied to electron occupation with a width of 0.05 eV. All relaxations and analysis, unless otherwise specified, were conducted using the BEEF-vdW XC.

% Electrochemistry
\subsection{Reaction Mechanism for NRR}
The overall reaction for NRR can be summarized as:
\begin{equation}
    \ce{N2} + 6(\ce{H+} + \ce{e-}) \rightarrow{} 2\ce{NH3}
\end{equation}
But, this can occur via many different possible mechanisms which are usually classified as either dissociative or associative. Since pure g-\ce{C3N4} is electrochemically inert \cite{Ma2014GraphiticElectrocatalysts,Zheng2014HydrogenElectrocatalyst}, and the dissociative mechanism requires two active sites, only the associative mechanisms are considered in this work. The associative mechanisms can be further subdivided into the distal, alternating, and enzymatic pathways. For all reactions below, $ ^*$ indicates adsorption. The distal mechanism proceeds as described in the following expressions:

% Distal
\begin{subequations}
\label{eq:dist}
\begin{gather}
     \ce{N2} + 6(\ce{H+} + \ce{e-}) + * \rightarrow{} \ce{NNH}^* + 5(\ce{H+} + \ce{e-})\\
     \ce{NNH}^* + 5(\ce{H+} + \ce{e-}) \rightarrow{} \ce{NNH2}^* + 4(\ce{H+} + \ce{e-})\\
     \ce{NNH2}^* + 4(\ce{H+} + \ce{e-}) \rightarrow{} \ce{N}^* + 3(\ce{H+} + \ce{e-}) + \ce{NH3}\\
     \ce{N}^* + 3(\ce{H+} + \ce{e-}) \rightarrow{} \ce{NH}^* + 2(\ce{H+} + \ce{e-})\\
     \ce{NH}^* + 2(\ce{H+} + \ce{e-}) \rightarrow{} \ce{NH2}^* + (\ce{H+} + \ce{e-})\\
     \ce{NH2}^* + (\ce{H+} + \ce{e-}) \rightarrow{} \ce{NH3} + *
\end{gather}
\end{subequations}
In this mechanism, protons and electrons are donated until the outer N is fully saturated and released as \ce{NH3} before the anchoring N begins to be protonated. On the other hand, the alternating mechanism rotates between giving protons and electrons to the outer N and the anchoring N. This is summarized as follows:

% Alternating
\begin{subequations}
\label{eq:alt}
\begin{gather}
   \ce{N2} + 6(\ce{H+} + \ce{e-}) + * \rightarrow{} \ce{NNH}^* + 5(\ce{H+} + \ce{e-})\\
   \ce{NNH}^* + 5(\ce{H+} + \ce{e-}) \rightarrow{} \ce{NHNH}^* + 4(\ce{H+} + \ce{e-})\\
   \ce{NHNH}^* + 4(\ce{H+} + \ce{e-}) \rightarrow{} \ce{NHNH2}^* + 3(\ce{H+} + \ce{e-})\\
   \ce{NHNH2}^* + 3(\ce{H+} + \ce{e-}) \rightarrow{} \ce{NH2NH2}^* + 2(\ce{H+} + \ce{e-})\\
   \ce{NH2NH2}^* + 2(\ce{H+} + \ce{e-}) \rightarrow{} \ce{NH2}^* + (\ce{H+} + \ce{e-}) +\ce{NH3}\\
   \ce{NH2}^* + (\ce{H+} + \ce{e-}) \rightarrow{} \ce{NH3} + *
\end{gather}
\end{subequations}
For both of the mechanisms described above, the bond between the nitrogens are perpendicular to the basal plane. The enzymatic mechanism however is analogous to the alternating mechanism except with the N-N bond parallel to the lattice plane. The competing HER reaction is a 2\ce{e-} process which can be described as \cite{Nrskov2005TrendsEvolution}:

% HER
\begin{subequations}
\begin{gather}
    2(\ce{H+} + \ce{e-}) + * \rightarrow \textrm{H}^* + (\ce{H+} + \ce{e-})\\
    \textrm{H}^* + (\ce{H+} + \ce{e-}) \rightarrow \ce{H2} + *
\end{gather}
\end{subequations}

For all adsorption energy and reaction energy calculations the reference was set to gas-phase \ce{N2} and \ce{H2}. The reference electrode was set to the computational hydrogen electrode where $ \frac{1}{2}G_{\ce{H2}} \rightleftharpoons G_{\ce{H+}} + G_{\ce{e-}}$ is at equilibrium, and a pH of 0. Adsorption energies for a given NRR intermediate were calculated as follows:
\begin{equation}
    \Delta E_{\textrm{N}_x\textrm{H}_y^*} = E_{\textrm{N}_x\textrm{H}_y^*} - \frac{x}{2}E_{\ce{N2}} - \frac{y}{2}E_{\ce{H2}}-E_*
\end{equation}
To calculate the gibbs adsorption energy at each step of the reaction the following expression was used:

\begin{equation}
    \Delta G_{\textrm{N}_x\textrm{H}_y^* + z\ce{NH3}} = G_{\textrm{N}_x\textrm{H}_y^*} + zG_{\ce{NH3}} - \frac{x}{2}G_{\ce{N2}} - \frac{y}{2}G_{\ce{H2}}-G_*
\end{equation}
For each of these energies their vibrational contributions (zero-point energy and entropy) uses the harmonic approximation at a temperature of 300 K. The catalyst single atom is held fixed for computational convenience, as allowing it to vibrate is observed to have a negligible effect on the zero-point energy and entropy values.

% BEEF
\subsection{Uncertainty Quantification}
To quantify the uncertainty, ensembles of energies are obtained using the BEEF-vdW XC functional which takes the following form \cite{Wellendorff2012DensityEstimation}:

\begin{equation}
    E_{\textrm{XC}}^{\textrm{BEEF-vdW}} = \sum_m a_m E_m^{\textrm{GGA-x}} + \alpha_c E^{\textrm{LDA-c}} + (1-\alpha_c)E^{\textrm{PBE-c}} + E^{\textrm{nl-c}}
\end{equation}

Van der Waals contributions to the energy are accounted for in $ E^{\textrm{nl-c}}$, a vdW-DF2 nonlocal correlation \cite{Wellendorff2011OnFunctionals}. The GGA exchange energy is projected onto Legendre polynomials giving the parameters $a_m$. Trade-off between the Perdew-Burke-Ernzerhof (PBE) correlation \cite{perdew1996generalized} and Perdew-Wang LDA correlation \cite{PhysRevB.45.13244} yields an additional parameter $\alpha_c$. Optimal parameters were obtained in the original formulation of this functional by fitting to a variety of condensed matter and chemical systems including molecular chemisorption on solid surfaces, noncovalent interactions, and molecular reaction energies. Using a Bayesian approach, these datasets also allowed for the generation of a posterior probability distribution for the parameters as $P(\textbf{a}| \theta; D) \sim e^{-C(\textbf{a})/ \tau}$. Here, $\textbf{a}$ is the set of parameters, $ \theta$ is the model, $ D$ is the data, $C(\textbf{a})$ is the cost function, and $\tau$ is a cost temperature. After using the optimized parameters for SCF, the parameter space is sampled from this posterior distribution to give an ensemble of energies. In this work we use an ensemble of 2000 XC functionals. Integration of this functional into existing DFT workflows is relatively straightforward with little sacrificed in terms of computational cost.

% Confidence 
From this ensemble, confidence values associated with important mechanistic and catalytic properties are calculated. The first confidence value obtained is associated with the confidence that the distal mechanism will proceed on a given surface:

\begin{equation}
\label{eqn:c_NNH2}
    c_{\ce{NNH2}}=\frac{1}{N_{ens}} \sum_{n=1}^{N_{ens}} \Theta (\Delta G_{\ce{NHNH}^*}^n-\Delta G_{\ce{NNH2}^*}^n)
\end{equation}
where $ N_{ens}$ is the number of XC functionals in the BEEF-vdW ensemble, $\Theta(x)$ is the Heaviside step function, $ \Delta G_{\ce{NHNH}^*(\ce{NNH2}^*)}^n$ is the reaction energy for \ce{NHNH}$^*$ (\ce{NNH2}$^*$) using the $n$-th XC functional. Both the distal and alternating mechanisms begin by forming \ce{NNH}$^*$. After this step the mechanisms diverge, and there is an uncertainty associated with path selection. The confidence value described in eq.\ref{eqn:c_NNH2} addresses this by quantifying the confidence as to whether the distal mechanism would be favourable over the alternating mechanism. 

Another source of uncertainty in these systems is from the predicted limiting potentials for these systems. For this reaction which contains uphill steps, the limiting potential is $ \textrm{U}_\textrm{L} = -\frac{1}{e}\textrm{max}\{\Delta G_1, \ldots , \Delta G_6\}$ with $ \Delta G_i$ being the free energy change for the $i$-th step. Propagating the energy ensembles in turn yields an ensemble of $ \textrm{U}_\textrm{L}$ values. The reaction step which determines this limiting potential is generally referred to as the potential determining step (PDS). Confidence in predicting that the PDS is step $i$ is computed using:

\begin{equation}
\label{eqn:c_pds}
  c_{PDS=i}=\frac{1}{N_{ens}} \sum_{n=1}^{N_{ens}} \delta_{m_{\textrm{pred}}^n,i}  
\end{equation}
where $m_{\textrm{pred}}^i$ is a classifier that identifies the most thermodynamically likely PDS for the $n$-th XC functional and $ \delta_{k,k'}$ is the Kronecker delta function. 

The final area of uncertainty when considering these systems for NRR is that of selectivity against HER. There is uncertainty over whether the NRR limiting potential is more negative then the HER limiting potential. If the limiting potential of NRR is less negative than HER, then NRR could be activated without also activating HER. The confidence that the limiting potential of NRR will not activate HER as well is described as follows:

\begin{equation}\label{eqn:c_HER}
    c_{\text{\sout{HER}}}^{\textrm{U}_\textrm{L}} = \frac{1}{N_{ens}} \sum_{n=1}^{N_{ens}} \Theta (\textrm{U}_\textrm{L,\ce{NRR}}^n-\textrm{U}_\textrm{L,\ce{HER}}^n)
\end{equation}
where $\textrm{U}_\textrm{L,\ce{NRR/HER}}^n$ is the limiting potential for NRR/HER corresponding to the free energy landscape generated by the $n$-th XC functional. Taking this a step further, the selectivity will also be influenced by whether H$^*$ or NNH$^*$ is more thermodynamically favourable to form on the surface. Thus, a confidence in whether NRR can be activated without activating HER while NNH$^*$ is also the more favorable adsorbate can be calculated as:
\begin{equation}\label{eqn:c_NRR}
    c_{\textrm{NRR}} = \frac{1}{N_{ens}} \sum_{n=1}^{N_{ens}} [\Theta (\Delta G_{\ce{NNH^*}}^n-\Delta G_{\ce{H^*}}^n)][\Theta (\textrm{U}_\textrm{L,\ce{NRR}}^n-\textrm{U}_\textrm{L,\ce{HER}}^n)]
\end{equation}
Put differently, $ c_{\textrm{NRR}}$ describes the confidence that the system will be able to promote NRR where there is no competition with HER. It is important to distinguish this quantity from the more traditional view of selectivity which is the percentage of the reactants proceeding via NRR instead of HER.

% $$ \Delta E_{(\ce{N_xH_y^*}+z\ce{NH3}(g))}=E_{\ce{N_xH_y^*}}+zE_{\ce{NH3}(g)}-E_{*}-\frac{x}{2}E_{\ce{N2}(g)}-\frac{y}{2}E_{\ce{H2}(g)}$$

% $$ \Delta F^{vib}_{\ce{N_xH_y^*}}=ZPE_{\ce{N_xH_y^*}}-TS_{\ce{N_xH_y^*}}-\frac{x}{2}(ZPE_{\ce{N2}(g)}-TS_{\ce{N2}(g)})-\frac{y}{2}(ZPE_{\ce{H2}(g)}-TS_{\ce{H2}(g)})$$

% $$ \Delta G_{(\ce{N_xH_y^*}+z\ce{NH3}(g))} = \Delta E_{(\ce{N_xH_y^*}+z\ce{NH3}(g))} + \Delta F^{vib}_{\ce{N_xH_y^*}} + zF^{vib}_{\ce{NH3}(g)} +eU_{SHE} $$

\section{Results and discussion}

%%%% Benchmarking -> Free Energy Landscapes -> Bifurcation -> PDS -> Scaling -> Selectivity %%%%%
\subsection{Benchmarking Uncertainty Estimates} \label{sec:bench}
We first obtain the single atom systems by relaxing single atoms of Pt, Fe, and Ru onto the clean g-\ce{C3N4} surface. It is observed that for Fe adsorption, the cavity was the thermodynamically optimal adsorption position. Based on this, we similarly relax both Pt and Ru within the cavity to obtain their optimal configurations.   

\begin{figure}
%  As well as the standard float types \texttt{table}\\
%  and \texttt{figure}, the class also recognises\\
%  \texttt{scheme}, \texttt{chart} and \texttt{graph}.
  \includegraphics[scale=0.55]{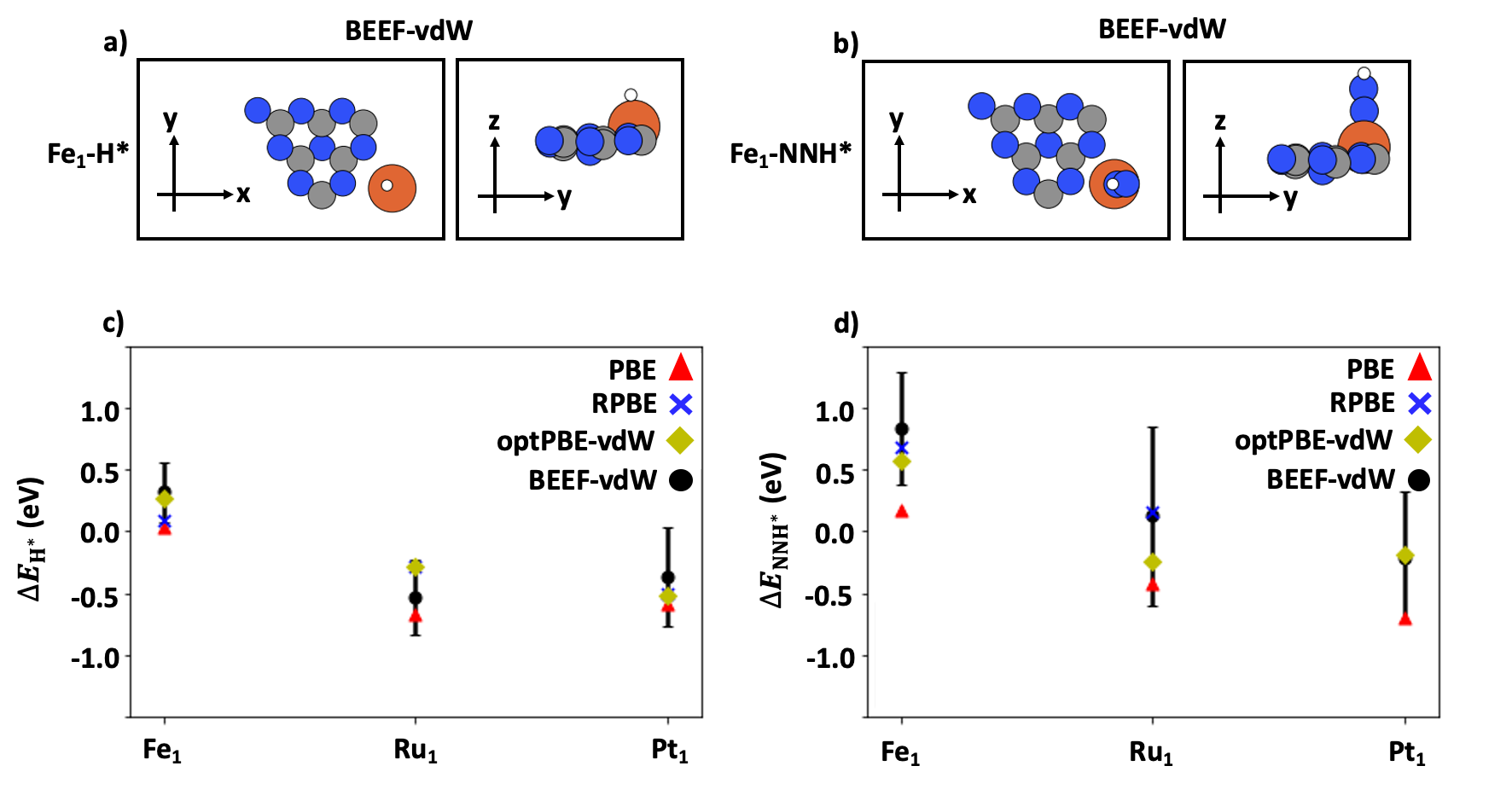}
  \caption{Relaxed geometrical configuration of a) H* and b) NNH* on Fe$_1$ using the optimal BEEF-vdW XC (Color code: blue= N, grey = C, Fe = orange, white = H). Adsorption energies of c) H* and d) NNH* implementing PBE, RPBE, optPBE-vdW, and BEEF-vdW exchange correlation functionals. The errorbars are $\pm 2\sigma^i_{\ce{H}^*/\ce{NNH}^*}$ where $\sigma^i_{\ce{H}^*/\ce{NNH}^*}$ is the standard deviation of the BEEF-vdW ensembles on the $i$-th system for HER/NRR}
  \label{fgr:bench}
\end{figure}

To solidify the approach of uncertainty quantification with BEEF-vdW, it is important to benchmark the ensemble spread against the spread from selected popular XC functionals. As representative adsorbates, we relax both H* and NNH* onto each of the three systems using BEEF-vdW, PBE, RPBE \cite{Hammer1999ImprovedFunctionals}, and optPBE-vdW \cite{Klimes2010ChemicalFunctional} XCs. In figures \ref{fgr:bench}a,b the relaxed geometries on Fe$_1$ from the optimal BEEF-vdW parameters are shown. It is observed that the presence of the adsorbates on the surfaces brought the single atom (SA) out of plane. This indicates the adsorbate modifies the interaction with the substrate, and highlights the dynamism of these systems as they undergo a given reaction. From each of these different combinations, adsorption energies of the representative adsorbates are calculated and compared (Fig. \ref{fgr:bench} c,d). The astute reader may notice that for Pt$_1$ does not have an adsorption energy reported for NNH* using the RPBE XC. This is because for this XC, NNH*  was observed to desorb from the surface during relaxation. However, this was the only system observed to demonstrate this behavior.

As an estimate of the uncertainty,the standard deviations of the BEEF-vdW ensembles, $\sigma^i_{\ce{H}^*/\ce{NNH}^*}$, is calculated on the $i$-th system. We justify treating these distributions as normal by calculating their skewness and kurtosis, and verified that all systems have values close to 0 and 3 respectively (Figs. S1,S2). Treating the distributions in this way, the uncertainty presents itself in the figures as the error bars of $\pm 2\sigma^i_{\ce{H}^*/\ce{NNH}^*}$. Almost all XC functionals of the ensemble predict adsorption energies fall within the range of the errorbars. This indicates that the ensemble is able to accurately reflect the sensitivity of the adsorption energies on the selected XC functional, and provides a systematic method to quantifying the uncertainty for these SA systems. 

\subsection{Reaction Mechanism Uncertainty}
When considering the associative reaction mechanisms (distal, alternating, and enzymatic), all three nucleate from adsorbed \ce{N2} on a clean catalyst surface which is subsequently protonated. However, the distinguishing feature of the enzymatic mechanism is that the \ce{N2} adsorbs horizontally onto the surface, whereas the distal and alternating mechanisms feature vertical adsorption. Thus, we adsorbed \ce{N2} in both a vertical and horizontal configuration onto each of the SA systems. All vertical configurations are found to be thermodynamically favorable with adsorption energies of -0.63 eV on Fe$_1$, -0.89 eV on Ru$_1$, and -1.33 eV on Pt$_1$. Therefore, this configuration is plausible on all three systems. On the other hand, the horizontal orientation of \ce{N2}* resulted in weaker binding in all cases of -0.20 eV, -0.37 eV, and -0.06 eV  on Fe$_1$, Ru$_1$, and Pt$_1$ respectively. However, protonation of these horizontal geometries leads to relaxation into vertical configurations on all systems. Therefore, we do not consider the enzymatic pathway, and only investigate the distal and alternating mechanisms. 

\begin{figure}
  \includegraphics[scale=0.575]{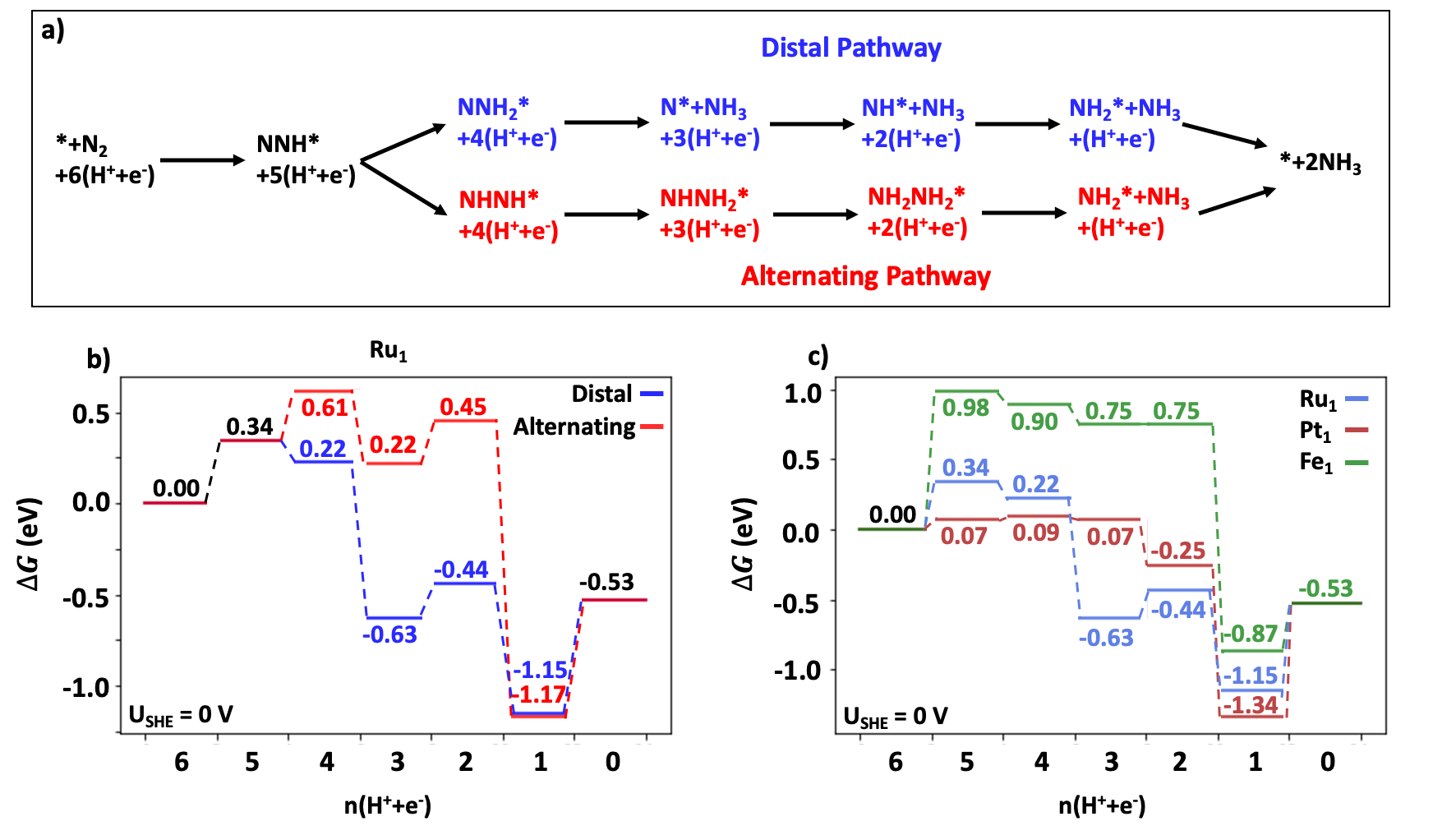}
  \caption{a) Reaction scheme for the distal and alternating mechanisms of ammonia synthesis b) Free energy landscape of both the distal and alternating mechanisms on the Ru$_1$ system at U$_{\textrm{SHE}}=0$ c) Free energy landscapes of the thermodynamically predicted dominant mechanisms on Ru$_1$, Pt$_1$, and Fe$_1$}
  \label{fgr:fed}
\end{figure}

Both mechanisms begin with NNH* formation before diverging into the distal and alternating paths which recombine at the final step with two units of ammonia released along the way (Fig. \ref{fgr:fed}a). Comparing these two paths the intermediates are quite varied and contain rich surface chemistries. Along the distal pathway, with ammonia being emitted at the halfway point, its adsorbates tend to contain less atoms than in the alternating pathway which does not release any ammonia until the last couple steps. It is important to understand which of the mechanisms is most likely to occur. Divergence in the reaction scheme results in a fork in the free energy landscape, as illustrated in figure \ref{fgr:fed}b using Ru$_1$ as an example. Similar plots for Fe$_1$ and Pt$_1$ are given in \textbf{Fig S3}. These landscapes are at $U=0$ relative to the standard hydrogen electrode (SHE). The contrasting energetics along the two mechanisms highlights the importance of quantifying confidence in which mechanism is favored. Traditionally, prediction of the favored path is based on whether \ce{NNH2}* or \ce{NHNH}* is thermodynamically favorable from NNH*. Using this approach, the traditionally predicted dominant mechanisms are shown in Fig. \ref{fgr:fed}c. Here, both Pt$_1$ and Fe$_1$ are expected to undergo an alternating mechanism, whereas Ru$_1$ is expected to undergo a distal mechanism. The importance of this is that in order to tune the free energy landscapes shown to maximize catalytic activity, the design principles for Ru$_1$ could be inherently different than the others based on this mechanistic distinction. Therefore, it is important to attribute a confidence value to what we would expect to be the dominant reaction mechanism before any broad claims could be justifiably made.

\begin{figure}
  \includegraphics[scale=0.625]{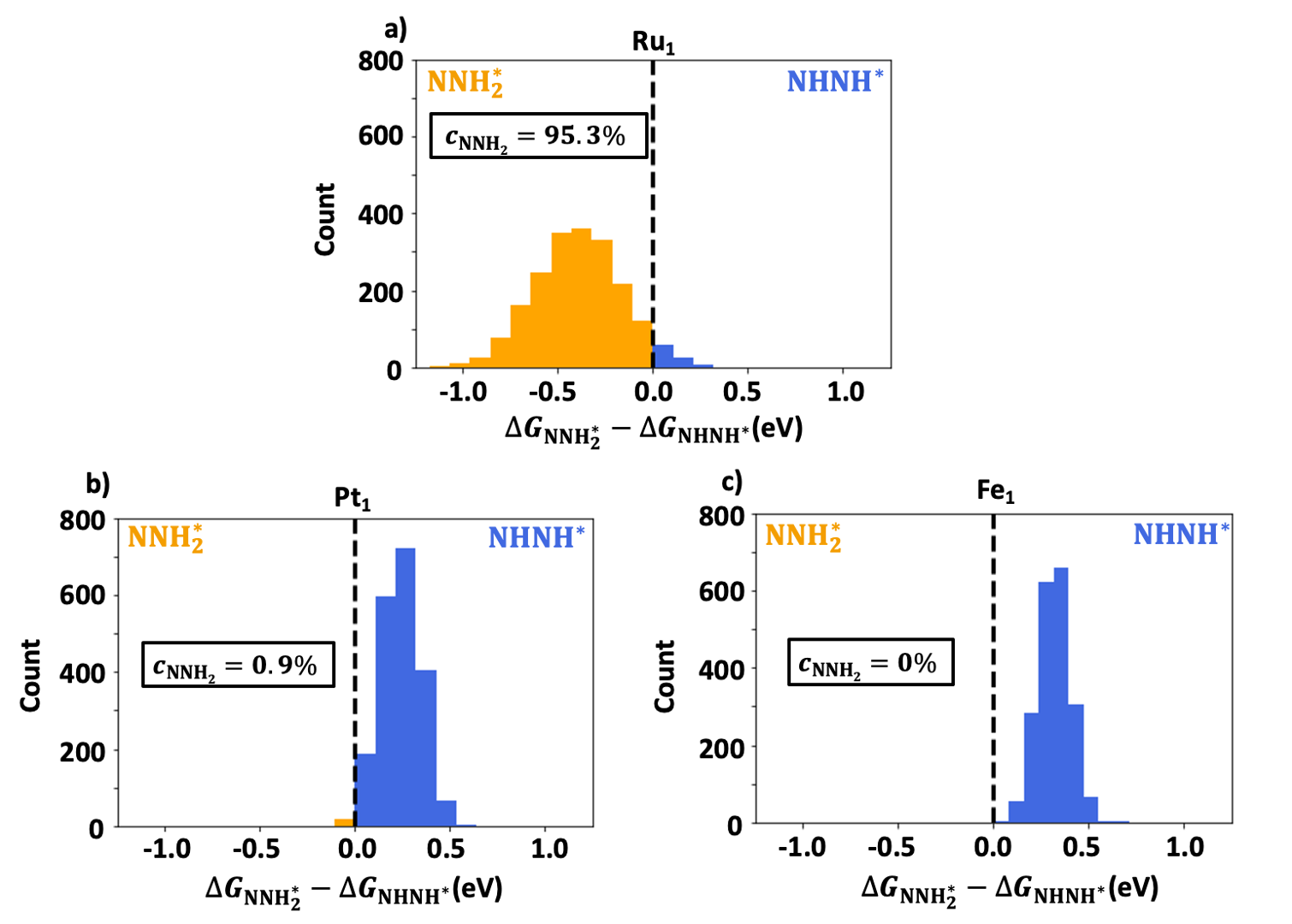}
  \caption{Histograms of the ensemble-wise difference between $\Delta G_{\ce{NNH2}^*}$ and $\Delta G_{\ce{NHNH}^*}$ on a) Ru$_1$ b) Fe$_1$ c) Pt$_1$. A negative difference indicates that the functional prefers a distal mechanism, and a positive difference indicates a preference towards the alternating mechanism. The inset confidence values are calculated using eq. \ref{eqn:c_NNH2} and indicate the fraction of functionals within the ensemble that favour \ce{NNH2}* formation} 
  \label{fgr:bif}
\end{figure}

For each adsorption energy of \ce{NNH2}* and \ce{NHNH}* in the ensemble, the difference is taken to obtain a histogram of dominant pathway predictions (Fig. \ref{fgr:bif}). Additionally, eq. \ref{eqn:c_NNH2} is used to obtain confidence values that the distal mechanism would be preferred. In the case of Ru$_1$ (Fig. \ref{fgr:bif}a), it is observed that most of the ensembles are in favor of \ce{NNH2}* formation opposed to \ce{NHNH}*, with a confidence value of 95.3 \%. However, it must be noted that there is a nonzero number of ensembles that prefer \ce{NHNH}* formation. Therefore, while there is a high confidence in the reaction proceeding via the distal mechanism, the alternating mechanism cannot be fully ruled out. Similarly for Pt$_1$ (Fig. \ref{fgr:bif}b), while the majority of functionals indicate \ce{NHNH}* being preferred, there is a nonzero number of functionals that suggest the distal mechanism could occur with a confidence of 0.9 \%. Therefore, in this case the distal mechanism cannot be entirely ruled out either. However, for the Fe$_1$ system, all of the functionals favor \ce{NHNH}*, and thus there is a high degree of confidence that the reaction will proceed exclusively via the alternating mechanism. In summary, the dominant mechanism varies on a case by case basis, and the degree of confidence with each mechanism also fluctuates. Therefore, the most likely mechanism to occur should be viewed in a probabilistic manner based upon confidence values obtained by the BEEF-vdW ensemble.

\subsection{Potential Determining Step Uncertainty}

\begin{figure}
  \includegraphics[scale=0.7]{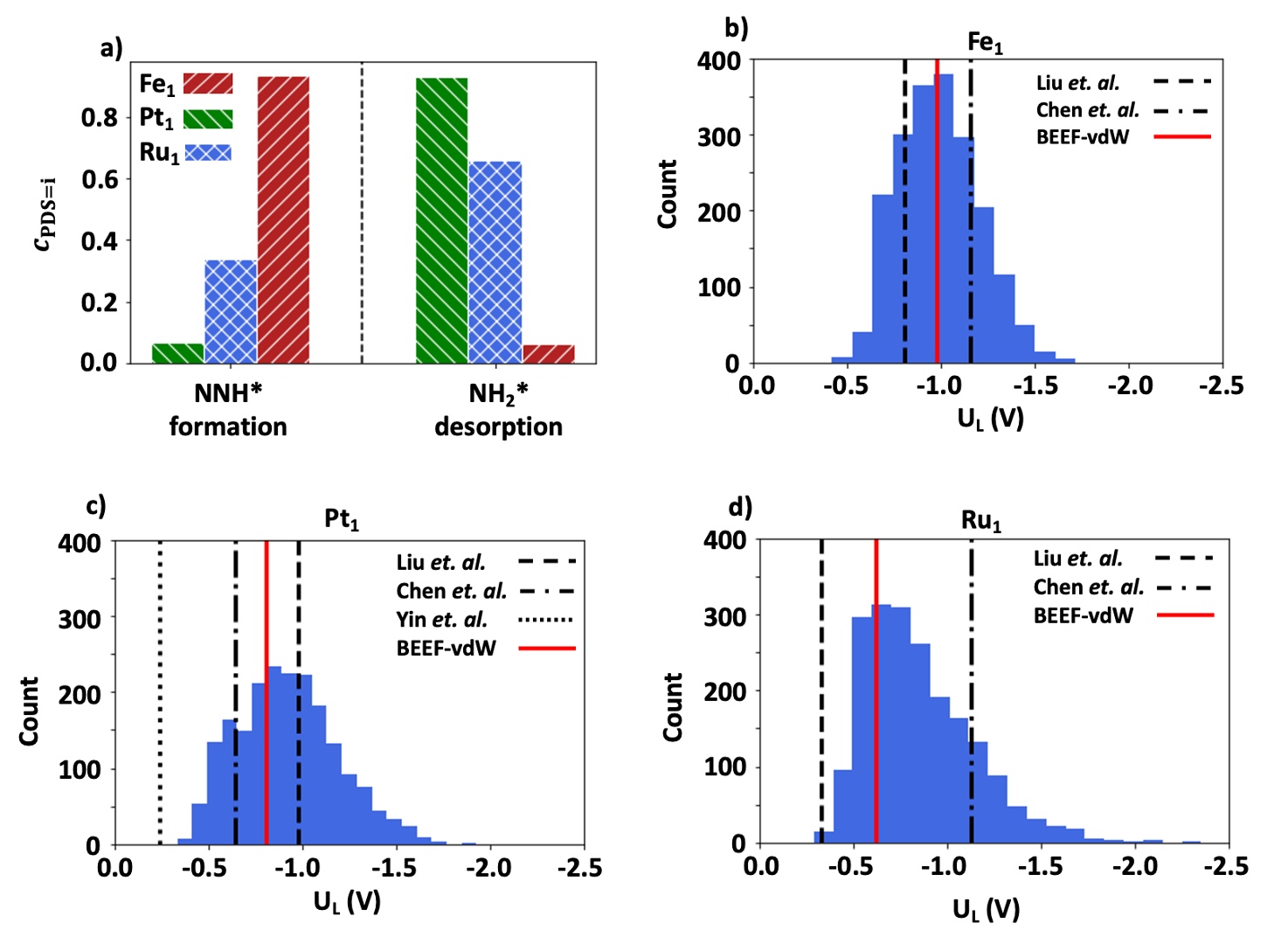}
  \caption{a) Confidence values of the potential determining step for each of the systems. The observed give and take between the confidence of these steps implies scaling. There is a 0.1 \% predicted confidence the PDS on Ru$_1$ was the protonation of N* which is not included. Distributions of the limiting potentials associated with b) Fe$_1$ c) Pt$_1$ d) Ru$_1$. Skewness for each of these distributions is calculated to be 0.17, 0.35, and 0.96 for Fe$_1$, Pt$_1$, and Ru$_1$, respectively. The dashed lines are U$_\textrm{L}$ values obtained from literature \cite{Yin2019Pt-embeddedSynthesis,Liu2019BuildingCatalysts,Chen2019ComputationalElectroreduction}. The red solid line is the limiting potential from the optimal BEEF-vdW XC functional}
  \label{fgr:pds}
\end{figure}

Shifting focus towards the predicted dominant mechanisms, the PDS determines the limiting potential and thus, to a first approximation, the electrocatalytic activity of a given system. Moreover, the nature of the step can lead to varying strategies when looking for future candidates.  For example, if the PDS was NNH* formation, then emphasis should be placed on tweaking the system to increase NNH* binding via methods such as tuning the local coordination environment \cite{Jung2020Atomic-levelProduction}. Thus, we quantify the confidence of the PDS for each of the systems using eq. \ref{eqn:c_pds}. These results are presented in Fig. \ref{fgr:pds}a, and it is observed that, for almost all functionals within the ensemble, the PDS is either NNH* formation or \ce{NH2}* desorption to form \ce{NH3}. In this figure, we can observe the two steps trade-off in confidence, i.e. higher confidence in one leads to lower confidence in the other. This provides strong evidence to the importance of these two steps towards activity, as proposed in previous studies \cite{Montoya2015TheRelations} on bulk metals. Here, it is observed that there is a high confidence that the PDS on Fe$_1$ is NNH* formation and \ce{NH2}* desorption on Pt$_1$. Therefore, strategies to improve these systems (ie. strengthening NNH* adsorption and weakening \ce{NH2} adsorption) can be confidently identified. However, on Ru$_1$ both PDS are relatively likely (with 0.34 and 0.66 confidence for NNH* and \ce{NH2}, respectively). Therefore, in this case of Ru$_1$ while weakening of \ce{NH2}* adsorption should be the focus for future improvement, the influence of NNH* formation cannot be ignored.

With each of the ensemble members yielding their own free energy landscape, a distribution of limiting potentials is extracted for each system (Fig. \ref{fgr:pds}b-d). Comparing these distributions to literature values \cite{Yin2019Pt-embeddedSynthesis,Liu2019BuildingCatalysts,Chen2019ComputationalElectroreduction}, it is observed that the distribution was able to bound most of them. This further solidifies this methodology of uncertainty quantification. Comparing the BEEF-vdW optimal limiting potentials, we find that they predict a hierarchy of Ru$_1 > \:$Pt$_1 > \:$Fe$_1$. However, further information can be extracted from the shape of the U$_\textrm{L}$ distributions. Calculating the skewness for each of the distributions we find their values to be 0.17 for Fe$_1$, 0.35 for Pt$_1$, and 0.96 for Ru$_1$. As the skewness increases with the optimal BEEF-vdW becoming less negative, an underlying maximum limiting potential is implied. Put differently, if the right side of Ru$_1$'s distribution is mirrored on the left, it would increase well into the positive regime. However, if a limit was to exist, a folding would occur causing a skewed distribution shape. Since scaling relations can cause the presence of activity volcanoes with a maximum achievable activity \cite{Krishnamurthy2018MaximalReactions,Kulkarni2018UnderstandingReaction}, these results indicate the presence of scaling within these systems. This will be discussed in greater detail in the following section. With Fe$_1$'s symmetry, we can conclude that it does not approach the peak U$_\textrm{L}$. For Pt$_1$ we predict the ensemble does have some folding from interacting with the volcano peak, while the large skewness of Ru$_1$ implies a high amount of interaction. Therefore, the distribution shape also suggests that Ru$_1$ has the best chance of reaching the maximum achievable activity.

\subsection{NRR Scaling Relations Uncertainty}\label{sec:scale}

\begin{figure}
  \includegraphics[scale=0.475]{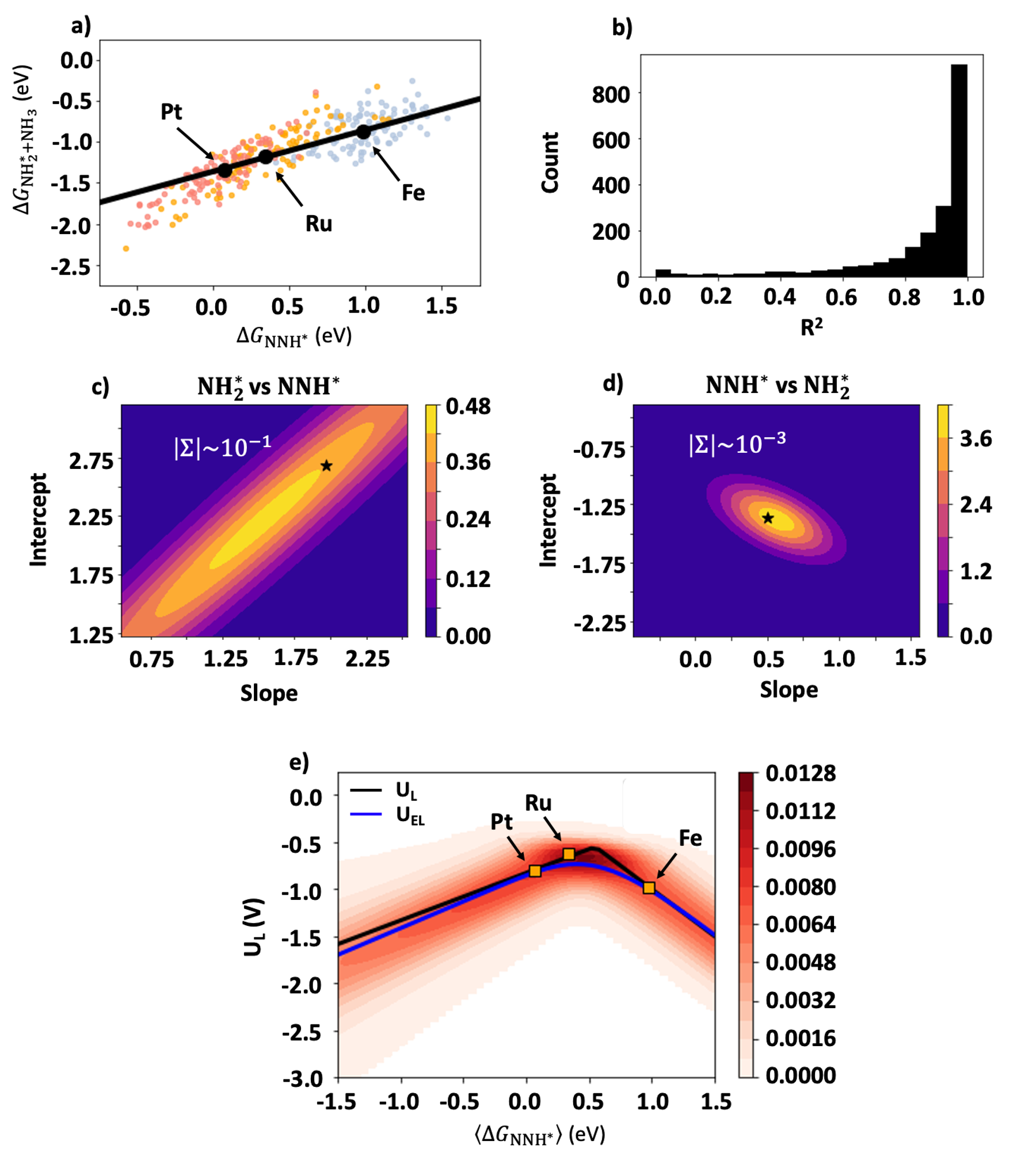}
  \caption{a) Scaling Relationship of $\Delta G_{\ce{NNH}^*}$ and $\Delta G_{\ce{NH2}^*+\ce{NH3}}$. Black dots are the optimal BEEF-vdW values, with the solid black line the corresponding linear fit. The red, orange, and blue dots correspond to a sampling of 100 XC functionals from the BEEF-vdW ensemble for Pt$_1$, Ru$_1$, and Fe$_1$, respectively. b) Distribution of correlation coefficients for each of the scaling relation fits in the ensemble. Probability density distribution in parameter space for when c) $\Delta G_{\ce{NH2}^*+\ce{NH3}}$ and d) $\Delta G_{\ce{NNH}^*}$ are the descriptors. The black stars correspond to the parameters from the optimal BEEF-vdW fit. e) Probabilistic activity volcano for NRR on these systems with the descriptor $\langle \Delta G_{\ce{NNH}^*} \rangle$. The solid black line is the limiting potential based on the optimal BEEF-vdW fitting parameters obtained from the sample. The solid blue line is the expected limiting potential which is the limiting potential weighted by the probability distribution. Orange squares are the BEEF-vdW optimal values}
  \label{fgr:scale_nnh_nh2}
\end{figure}

The presence of scaling relations between intermediate adsorption energies enforces a limit to the maximum achievable activity, and thus we explore scaling among NNH*, \ce{NH2}*, and N*. Previous investigations have reported scaling among these three intermediates \cite{Skulason2012AReduction,Liu2019BuildingCatalysts,Montoya2015TheRelations}, with particular emphasis placed on the former two due to their importance towards activity. Due to the observed give and take between $c_{PDS=i}$ among NNH* formation and \ce{NH2}* desorption, we first study the scaling between these two intermediates (Fig. \ref{fgr:scale_nnh_nh2}a). The solid black line is the linear fit obtained for the optimal BEEF-vdW values, illustrated as black dots, which all fall close to this line of best fit indicating strong scaling. Extending this analysis to include the uncertainty estimates of the BEEF-vdW ensembles, clusters of adsorption energies are obtained corresponding to each system. For every XC in the ensemble, there is one point within each of the system's clusters, thereby creating sets of three points. Therefore, for each of the 2000 XC functionals within the ensemble, a scaling relation can be extracted via an ordinary least squares fit creating an ensemble of scaling relations. It is observed that the clusters do follow a linear trend with each other, further implying the presence of strong scaling. To quantitatively assess the degree of scaling, we generate a histogram of the R$^2$ values for all 2000 scaling relations shown in Fig. \ref{fgr:scale_nnh_nh2}b. For the majority of the relations in the ensemble, the correlation coefficient is quite high, lying mainly around 1.0. Therefore, the adsorption energies for most of the XC functionals in the BEEF-vdW ensemble are quite linear. It is worth noting that for these fits no assumptions are made about the slope of the scaling relation nor the intercepts, allowing for a generalized ensemble of scaling relationships.

A prior work studied the systematic error present in scaling relations in oxygen reduction via BEEF-vdW \cite{Christensen2016FunctionalCatalysts,Deshpande2016QuantifyingReaction}. Here, we instead propagate the BEEF-vdW ensemble to obtain an ensemble of scaling relations, and study the distribution shape in fitting parameter space to obtain further fundamental insights. We are given the freedom to select either $\Delta G_{\ce{NNH}^*}$ or $\Delta G_{\ce{NH2}^*}$ as the descriptor for these scaling relations. By allowing both the fitting parameters to be completely unrestricted, the choice of descriptor significantly influences the distribution in parameter phase space. To gain insights into the best descriptor and describe the scaling between NNH* and \ce{NH2}*, we fit a multivariate normal to the parameter distribution of the scaling relation ensemble. In Fig. \ref{fgr:scale_nnh_nh2}c, we plot the fit using \ce{NH2}* as the signal and NNH* as the response. Since the distribution is on an angle with respect to the axes, these fitting parameters cannot be decoupled and must be considered in tandem. To gives an estimate of the relation's robustness towards XC selection, the determinant of the covariance matrix, $|\Sigma|$, is computed to be of order 10$^{-1}$. In comparison, the fit using NNH* as the signal shows a much more constricted shape (Fig. \ref{fgr:scale_nnh_nh2}d). There is still some diagonal behaviour in the distribution so this maintains the coupled nature of the parameters. $|\Sigma|$ for this distribution quantitatively highlights this compression by being of order 10$^{-3}$, 2 orders of magnitude smaller than when using \ce{NH2}* as the descriptor. By considering the spread of the distribution in parameter space, identification of the more appropriate signal for describing these scaling relations emerges. This is because a tighter distribution in phase space is an indication of greater robustness towards computational parameter selection. In this case, since the distribution using NNH* as the signal yielded a more narrow distribution, we suggest that it is the more suitable descriptor for describing this specific scaling relation.

Since N* has been used as a descriptor in other NRR studies \cite{Singh2019StrategiesSynthesis,Liu2019BuildingCatalysts}, its scaling strength with NNH* is evaluated (\textbf{Fig. S4a}). The adsorption energy clusters are observed to be relatively isolated from each other, and some ambiguity arises in terms of their scaling. While some of the functionals demonstrate a strong scaling correlation coefficient, the majority of the functionals favor weak correlation with the largest bin at an R$^2$ of 0 (Fig. S4b). Therefore, we conclude that the scaling exhibited between these intermediates is relatively weak. Applying the same methodology as above to gauge the spread in parameter space, N* is identified as the better descriptor with its determinant of the covariance matrix being three orders of magnitude smaller than when using NNH* as the signal (\textbf{Fig. S4c,d)}. Similarly, considering the scaling between N* and \ce{NH2}*, weak scaling is observed among the three systems (\textbf{Fig. S5a,b}). While N* is once again the more suitable descriptor (\textbf{Fig. S5c,d)}, it describes a weaker scaling phenomenon. Therefore, we focus on the stronger observed scaling between NNH* and \ce{NH2}*, which is more influential towards overall predicted performance. The method we present here is generalizable and presents a robust framework to systematically identify the best scaling descriptors.

As scaling relations form the backbone of activity volcanoes, they highlight the importance of properly identifying a descriptor which is invariant to choice of computational parameters. The fundamental principle behind activity volcanoes is that limiting potentials may be expressed in terms of a single descriptor through the use of scaling relations. This results in a Sabatier-type relationship with a maximum achievable activity allowed by scaling for a system corresponding to a given descriptor. In this case, since we have identified NNH* and \ce{NH2}* as the two intermediates with the most influence on predicted activity, and they demonstrate strong scaling to a high confidence, they are critical in determining the properties of the activity volcano. The limiting potential may be expressed in the form:

\begin{equation}\label{eqn:lim_pot_nrr}
   \textrm{U}_\textrm{L} = -\frac{1}{e}\textrm{min}(\Delta G_{\ce{NNH}^*},\Delta G_{2\ce{NH3}}-\Delta G_{\ce{NH2}^* + \ce{NH3}}) 
\end{equation}
From the presence of scaling between $\Delta G_{\ce{NNH}^*}$ and $\Delta G_{\ce{NH2}^* + \ce{NH3}}$, this expression can then be simplified to only depend on either $\Delta G_{\ce{NNH}^*}$ or $\Delta G_{\ce{NH2}^* + \ce{NH3}}$. We chose the former due to its stability in phase space. Therefore, for the $i$-th scaling relation the limiting potential is given by:

\begin{equation}\label{eq:volc}
    \textrm{U}_\textrm{L}^{(i)}(\Delta G_{\ce{NNH}^*}) = -\frac{1}{e}\textrm{min}(\Delta G_{\ce{NNH}^*},\Delta G_{2\ce{NH3}}-m^{(i)}\Delta G_{\ce{NNH}^*}-b^{(i)}) 
\end{equation}
where $m^{(i)}$ and $b^{(i)}$ are from the $i$-th member of the scaling ensemble. Stemming from Eq. \ref{eq:volc}, the path to creating an ensemble of activity volcanoes emerges; propagation of the scaling relations obtained in Fig. \ref{fgr:scale_nnh_nh2}a. This allows for a probabilistic investigation into the activity volcano following a similar approach we outlined previously \cite{Krishnamurthy2018MaximalReactions}, and is described in detail in the \textbf{Supporting Information}. Choosing the expectation value of a given ensemble $\Delta G_{\ce{NNH}^*}$ as the descriptor, we calculate the probabilistic activity volcano for NRR on these single atom systems (Fig. \ref{fgr:scale_nnh_nh2}e). The uncertainty in this activity volcano stems from two sources: i) spread in the combined distribution of $\Delta G_{\ce{NNH}^*}$ and ii) variability in the scaling relations of NNH* and \ce{NH2}*. Firstly, all three BEEF-vdW ensembles for $\Delta G_{\ce{NNH}^*}$ on each of the three systems are superimposed into a single distribution. The standard deviation, $\sigma_{\ce{NNH}}$, is then calculated for this combined distribution to give an uncertainty estimate in the descriptor. This provides the uncertainty estimate for the weaker binding leg of the volcano. Secondly, the ensemble of scaling relations creates another dimension of uncertainty which is the source of uncertainty in the stronger binding leg. Propagating the uncertainty allows us to calculate a conditional probability density function $p(\textrm{U}_\textrm{L}|\langle \Delta G_{\ce{NNH}^*}\rangle)$ which highlights the most probable regions of limiting potential values given an ensemble average of $\Delta G_{\ce{NNH}^*}$ for an arbitrary system. This conditional probability density function is the contour in Fig. \ref{fgr:scale_nnh_nh2}e. The limiting potential found from the optimal BEEF-vdW fitting parameters is able to explain the investigated systems to a high degree. A similar quantity, the expected limiting potential, weights the limiting potential by its conditional probability as follows:

\begin{equation}
    \textrm{U}_\textrm{EL}(\langle \Delta G_{\ce{NNH}^*}\rangle) = \int_{-\infty}^{\textrm{U}_\textrm{L}^{\textrm{max}}} \textrm{U}_\textrm{L} \: p(\textrm{U}_\textrm{L}|\langle \Delta G_{\ce{NNH}^*}\rangle) \: \textrm{dU}_\textrm{L}
\end{equation}
where $\textrm{U}_\textrm{L}^{\textrm{max}}$ is the maximum observed limiting potential across all activity volcanoes in the ensemble. Near the peak of the volcano the expected limiting potential diverges, highlighting the limitations of the standard activity volcano alone in describing this regime. In corroboration with the skewness trend observed in the limiting potential distributions, Ru$_1$, which has the largest skewness, is closest to the peak where Fe$_1$ with the lowest skewness is furthest from the peak. Therefore, we conclude that the degree of skewness is a consequence of this imposed maximum activity of the activity volcano. In further agreement with the limiting potential distributions, Ru$_1$ has the highest potential for yielding the best NRR activity as the conditional probability for its descriptor is the most condensed in the region of peak activity. Additionally, the location of the BEEF-vdW optimal values on this volcano agree well with the computed activity volcano. This demonstrates the strength in the observed scaling relation, particularly as both Ru$_1$ and Fe$_1$ fall on the stronger binding leg which is described by this relation. Extracting design criteria from this volcano, the maximum activity on the $\textrm{U}_\textrm{L}$ volcano of -0.56 V lies at a $\langle \Delta G_{\ce{NNH}^*}\rangle$ of 0.52 eV. However, on the $\textrm{U}_\textrm{EL}$ curve the maximum activity of -0.75 lies at a $\langle \Delta G_{\ce{NNH}^*}\rangle$ of 0.40 eV. In comparison with Ru$_1$'s $\langle \Delta G_{\ce{NNH}^*}\rangle$ of 0.34 eV, it is extremely close to the peak of the $\textrm{U}_\textrm{EL}$ volcano. Therefore, in terms of looking for a material that excels at NRR activity, Ru$_1$ is a good place to start, and tweaking of this system to address selectivity is a promising avenue of exploration. In short, uncertainty estimates propagated through the scaling relations could explain activity behaviour and provide design criteria for within a given materials class.

% \begin{figure}
%   \includegraphics[scale=0.575]{scaling_nnh_n_updated.png}
%   \caption{a) Distribution of R$^2$ for $E_\textrm{NNH}$ vs $E_{\textrm{N}}$ (Inset: Ensemble of scaling relations; black line is fit of optimized BEEF-vdW functional); Fitting parameter space with b) NNH and c) N as the descriptor. Colour code: yellow = high concentration; purple = low concentration}
%   \label{fgr:scale_nnh_n}
% \end{figure}

\subsection{Selectivity Uncertainty}

\begin{figure}
  \includegraphics[scale=0.675]{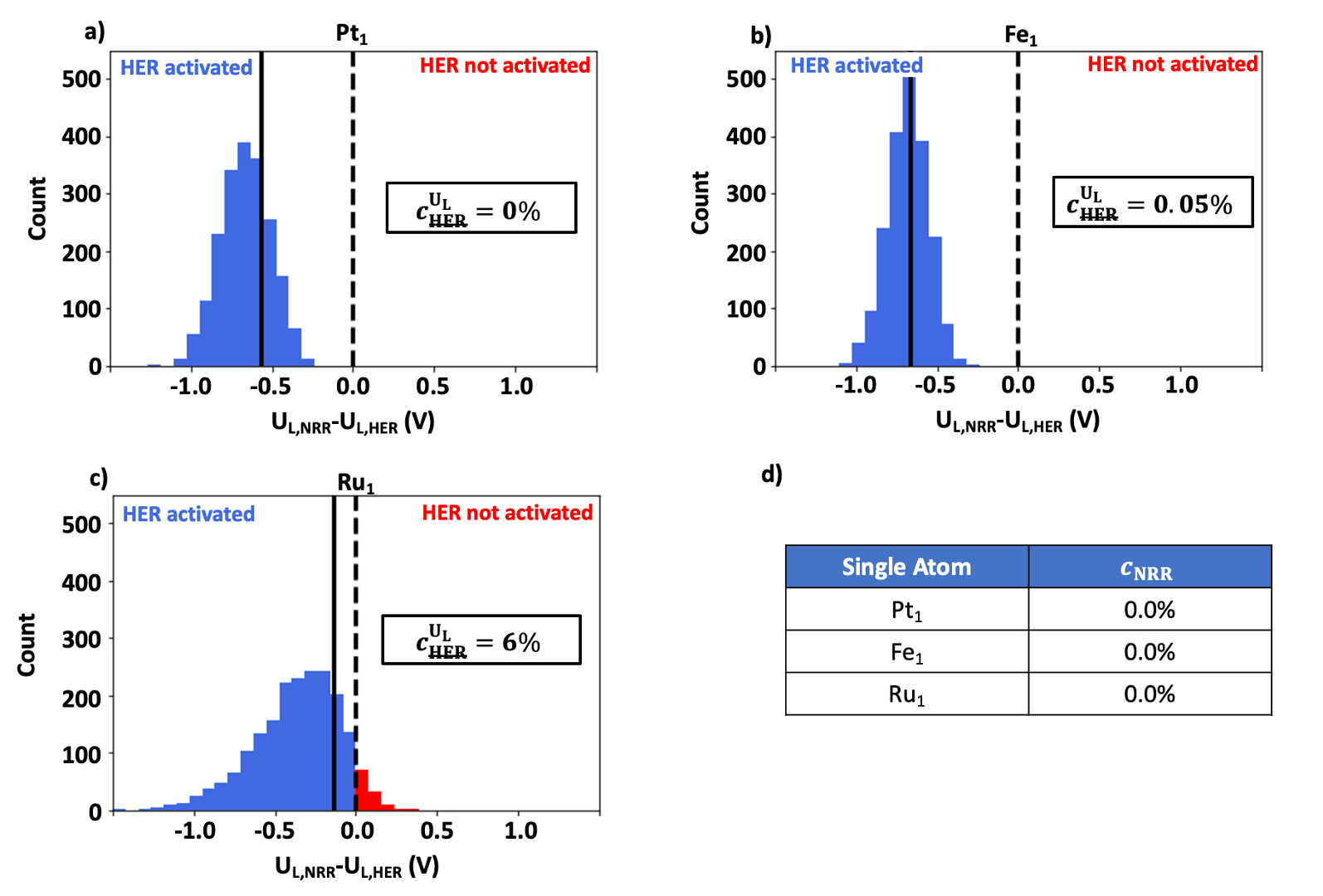}
  \caption{Histogram of the XCs that compare the limiting potentials of NRR and HER on a) Pt$_1$, b) Fe$_1$, and c) Ru$_1$. Functionals that have a positive difference indicate that HER could be suppressed. The solid black lines are the limiting potential difference from the optimal BEEF-vdW values. d) Confidence values that the competition between NRR and HER can be completely suppressed. None of the three systems indicate that they could completely block the presence of HER}
  \label{fgr:select}
\end{figure}

A major obstacle to the development of high performance NRR catalysts is the competition with the HER reaction \cite{Singh2019StrategiesSynthesis}. In this section, we outline a procedure to computationally evaluate an electrocatalyst's predicted selectivity capabilities with uncertainty estimation. To begin, we first compare the limiting potentials of NRR and HER. When the limiting potential of NRR falls below HER, then the potential required to activate NRR will also activate the parasitic HER reaction. Therefore, to improve NRR selectivity, we are striving for an NRR limiting potential that is less negative than HER. Thus, in Fig. \ref{fgr:select}a-c, we took an XC-wise difference of the limiting potentials of NRR and HER on each of the three systems. A larger fraction of functionals predicting a positive difference between these potentials implies a higher confidence for suppressing HER and better selectivity. On both Pt$_1$ and Fe$_1$, a small confidence of HER suppression $c_{\ce{HER}}^{\textrm{U}_\textrm{L}}$, calculated using eq. \ref{eqn:c_HER}, is observed. No XC functionals on Pt$_1$ predict HER suppression, where only 0.05 \% of the functionals on Fe$_1$ predict that HER could be suppressed. The optimal BEEF-vdW difference of the limiting potentials are also well into the negative region. On the other hand, 6 \% of the XC functionals predict that HER could be suppressed, the highest confidence of the three systems. Additionally, Ru$_1$'s optimal BEEF-vdW difference is closest to zero. Therefore, we predict Ru$_1$ to be the most likely to demonstrate improved selectivity. Expanding on this analysis we quantify the confidence that there will be no competition between NRR and HER, $c_{\textrm{NRR}}$. This is done using eq. \ref{eqn:c_NRR} which compares not only the limiting potential difference, but also whether \ce{NNH}* or \ce{H}* is favourable to adsorb on the surface. In other words, the latter term ensures that it is thermodynamically preferred for the surface to be covered by the nucleating \ce{NNH}* intermediate than \ce{H}*. Interestingly, all three systems exhibit $c_{\textrm{NRR}}$ values of 0, indicating that we predict none of the systems would be able to completely suppress HER, and at least some competition is expected to be present. Thus, further efforts into specifically addressing selectivity is necessary. 

\begin{figure}
  \includegraphics[scale=0.325]{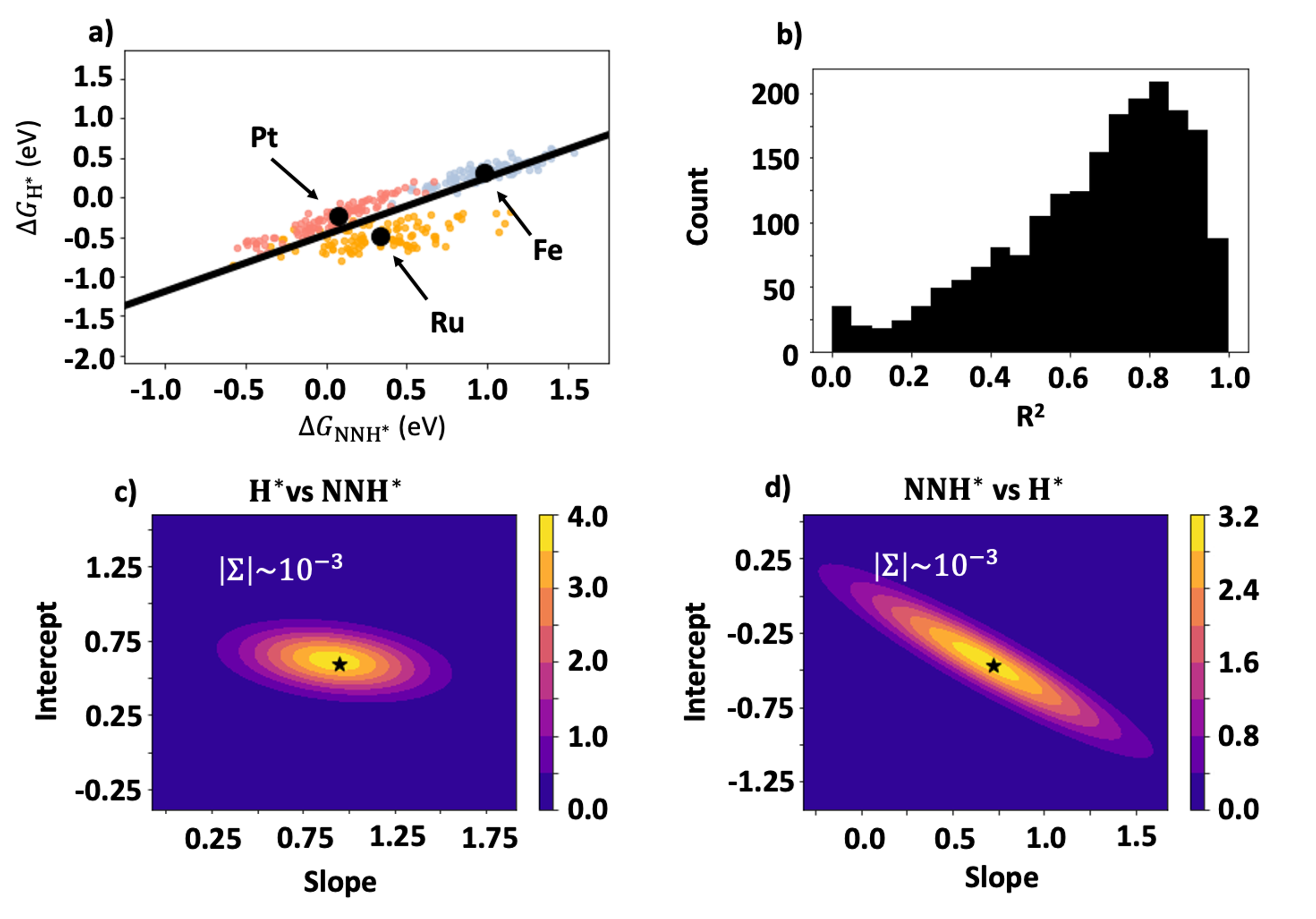}
  \caption{a) Scaling Relationship of $\Delta G_{\ce{NNH}^*}$ and $\Delta G_{\ce{H}^*}$. Black dots are the optimal BEEF-vdW values, with the solid black line the corresponding linear fit. The red, orange, and blue dots correspond to a sampling of 100 XC functionals from the BEEF-vdW ensemble for Pt$_1$, Ru$_1$, and Fe$_1$, respectively. b) Distribution of correlation coefficients for each of the scaling relation fits in the ensemble. Probability density distribution in parameter space for when c) $\Delta G_{\ce{H}^*}$ and d) $\Delta G_{\ce{NNH}^*}$ are the descriptors. The black stars correspond to the parameters from the optimal BEEF-vdW fit.}
  \label{fgr:select_scale}
\end{figure}

To further investigate the relationship between NRR and HER, we also study the scaling of the intermediate adsorption energies with the \ce{H}* adsorption energy. We draw particular attention to the scaling of $\Delta G_{\ce{NNH}^*}$ with $\Delta G_{\ce{H}^*}$ as the former was identified to be the best descriptor towards NRR activity. The scaling between these two quantities are plotted in Fig. \ref{fgr:select_scale}a and it is observed that the scaling does not appear as strong as that seen in Fig \ref{fgr:scale_nnh_nh2}a. This is then confirmed via the histogram analysis presented in Fig. \ref{fgr:select_scale}b where there is a larger spread in correlation coefficients. Therefore, while we can conclude that there is a considerable strength to the scaling from most of the XC's favoring high correlation coefficients, it is indeed not as strong as the \ce{NNH}* and \ce{NH2}* scaling. Applying a computational parameter stability analysis using both \ce{H}* and \ce{NNH}* as descriptors we observe that they are of both similar magnitude in terms of $|\Sigma|$ (Fig. \ref{fgr:select_scale}c,d). However, it is interesting to note that when using \ce{NNH}* as a descriptor, the distribution aligns much more diagonally in comparison to using \ce{H}* as a descriptor. Therefore, when using \ce{NNH}* as a signal for this scaling, the resulting computational parameters have a higher degree of coupling than when using \ce{H}*. This highlights the importance of descriptor selection, as in addition to affecting stability towards computational parameter selection, it also influences how the parameters interact with each other. Investigating the scaling relationship of \ce{N}* and \ce{NH2}* with \ce{H}* we observe interesting phenomena for both (\textbf{Fig. S6,S7}). For \ce{N}* and \ce{H}* we see similarly strong scaling as compared to using \ce{NNH}* which is in contrast to the weak scaling observed for \ce{N}* compared to other NRR intermediates. Additionally, when using \ce{N}* as the signal, the scaling is extremely resistant to computational parameter changes, with a $|\Sigma|$ of order $10^{-5}$. Alternatively, the scaling between \ce{NH2}* and \ce{H}* is observed to have widely varying scaling strength depending on the XC with many falling on either side of the spectrum. Moreover, the use of \ce{NH2}* as a descriptor is seen to be very sensitive to computational parameters with a $|\Sigma|$ of order $10^{-1}$.

\begin{figure}
  \includegraphics[scale=0.325]{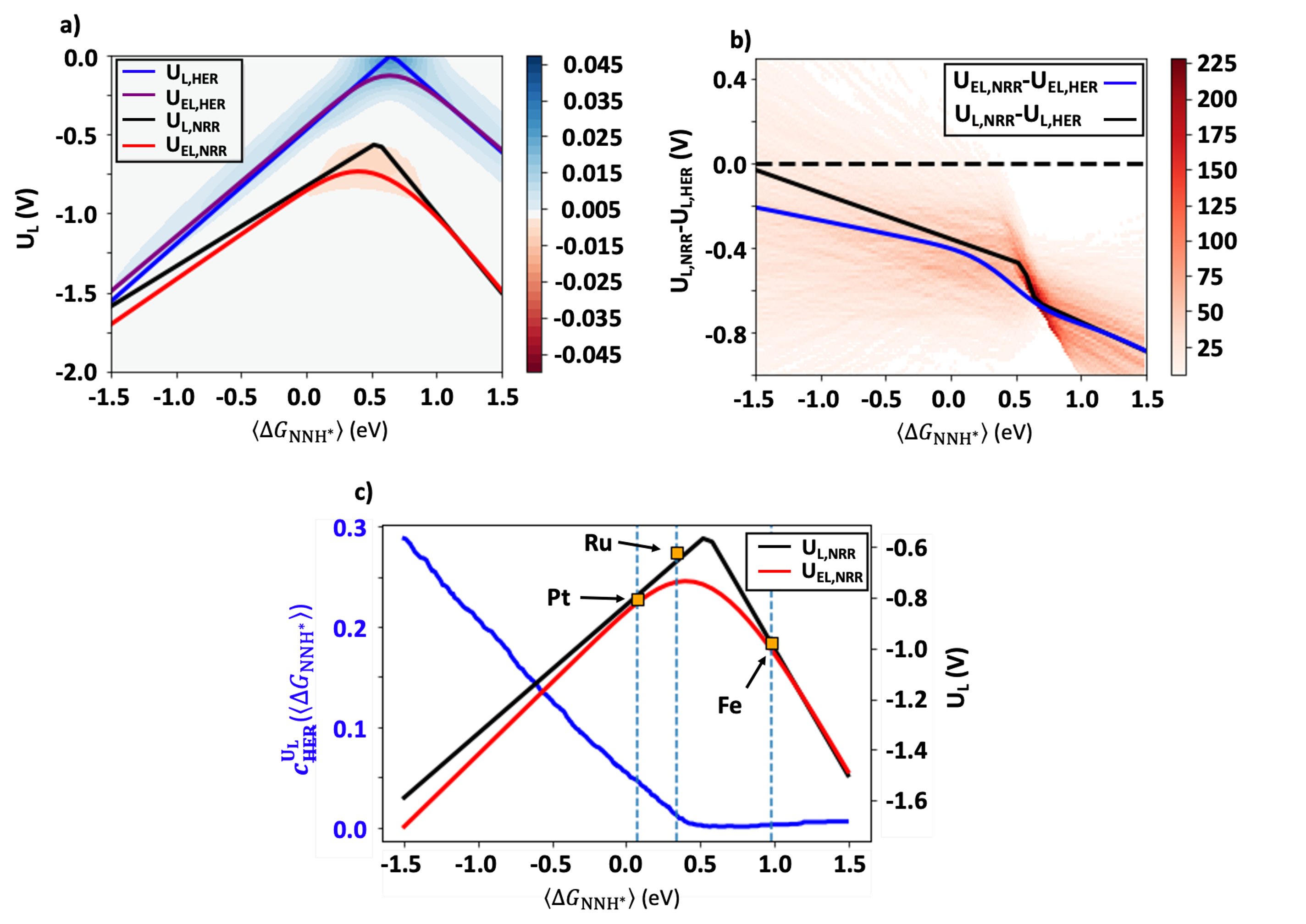}
  \caption{a) Combined probabilistic activity volcanoes for NRR and HER on the same domain of $\langle \Delta G_{\ce{NNH}^*}\rangle)$. The contour is the probability difference of $p(\textrm{U}_\textrm{L}^{\textrm{HER}}|\langle \Delta G_{\ce{NNH}^*}\rangle)$ and $p(\textrm{U}_\textrm{L}|\langle \Delta G_{\ce{NNH}^*}\rangle)$. Positive (negative) regions indicate the limiting potential given $\langle \Delta G_{\ce{NNH}^*}\rangle$ is more likely to be associated with HER (NRR). b) Limiting potential differences as a function of $\langle \Delta G_{\ce{NNH}^*}\rangle$. Contour is a 2D histogram of limiting potentials differences from propagation the volcano ensemble. Positive values indicate HER suppression. c) Confidence of HER suppression $c_{\text{\sout{HER}}}^{\textrm{U}_{\textrm{L}}}$ overlaid with the optimal BEEF-vdW NRR activity volcano and expected limiting potential curve. Descriptor value that optimizes both selectivity and activity is the intersection of these curves highlighted by the arrow}
  \label{fgr:select_volc}
\end{figure}

Due to the scaling of $\Delta G_{\ce{NNH}^*}$ with $\Delta G_{\ce{H}^*}$, we can then write an activity volcano for HER in terms of $\Delta G_{\ce{NNH}^*}$. The limiting potential for HER in general can be expressed as:

\begin{equation}
    \textrm{U}_\textrm{L}^{\textrm{HER}}(\Delta G_{\ce{H}^*}) = -\frac{1}{e}|\Delta G_{\ce{H}^*}|
\end{equation}
Substituting in for $\Delta G_{\ce{H}^*}$ using the scaling relations allows this to become a function of $\Delta G_{\ce{NNH}^*}$ and produce an ensemble of activity volcanoes where the $i$-th member can be expressed as:

\begin{equation}
    \textrm{U}_\textrm{L}^{(i),\textrm{HER}}(\Delta G_{\ce{NNH}^*})= \textrm{min}\left(-\frac{1}{e}(m^{(i)}\Delta G_{\ce{NNH}^*} + b^{(i)}), \frac{1}{e}(m^{(i)}\Delta G_{\ce{NNH}^*} + b^{(i)})\right)
\end{equation}
And, by applying the same procedure as for the NRR volcano, a conditional probability of $p(\textrm{U}_\textrm{L}^{\textrm{HER}}|\langle \Delta G_{\ce{NNH}^*}\rangle)$ can also be calculated for this reaction along with an expected limiting potential $\textrm{U}_\textrm{EL}^{\textrm{HER}}$ (\textbf{Fig. S8}). Moreover, making this change of variables allows for both the NRR and HER activity volcanoes to be compared on a shared domain (Fig. \ref{fgr:select_volc}a). On this combined plot we also evaluate the difference between $p(\textrm{U}_\textrm{L}^{\textrm{HER}}|\langle \Delta G_{\ce{NNH}^*}\rangle)$ and $p(\textrm{U}_\textrm{L}|\langle \Delta G_{\ce{NNH}^*}\rangle)$ which provides information on where the limiting potential is more likely to be associated with either NRR or HER. Firstly, it is observed that near the volcano peak the NRR volcano lay well below the HER volcano, and the probability differences are relatively concentrated. Therefore, near the peak we are most confident that the NRR peak lies below the HER peak, meaning that HER is activated and selectivity would be negatively impacted. Moving away from the peak the probability differences become negligible. On the HER volcano the probability differences become more dispersed than at the peak, but still relatively condensed. Along the NRR volcano however, the probability difference became smaller than the resolution of 0.5\%. In short, this means that while we have a reasonable degree of confidence as to where the limiting potential is more likely to be associated with HER when moving away from the peak, for NRR it becomes less clear. In terms of selectivity, this then indicates that moving away from the peak is necessary to increase the likelihood that the limiting potential of NRR could be such that HER is suppressed. It also becomes apparent from this volcano that the legs are not parallel. To explore this further, we take the difference of the limiting potentials as a function of $\langle \Delta G_{\ce{NNH}^*}\rangle$ (Fig. \ref{fgr:select_volc}b). On this plot, when the limiting potential difference becomes positive, it is predicted that NRR could be activated without activating HER. Propagating the scaling relations ensemble further, we generate histograms for each $\langle \Delta G_{\ce{NNH}^*}\rangle$ corresponding to the contour in the plot. Regions of high concentration indicate a higher confidence in the limiting potential difference given $\langle \Delta G_{\ce{NNH}^*}\rangle$. When $\langle \Delta G_{\ce{NNH}^*}\rangle$ is approximately 0.55-0.65 eV, which is near the volcano peaks, the range of limiting potential difference values is quite compressed. This supports the earlier observation that near the peaks there is less confidence in being able to suppress HER. Decreasing the binding strength of \ce{NNH}* on the surface (ie. increasing $\langle \Delta G_{\ce{NNH}^*}\rangle$) results in the limiting potential difference becoming more negative, thus straying from the ideal scenario of a positive difference in potential. On the other hand, stronger binding of \ce{NNH}* results in the difference approaching zero. Moreover, the spread in the differences widens on this side of the domain, indicating that some of the XCs predict a positive difference, and there becomes a nonzero probability of HER suppression. This behavior illuminates a selectivity-activity trade-off for these systems. As $\langle \Delta G_{\ce{NNH}^*}\rangle)$ approaches the volcano peak, the selectivity is likely to worsen. Moving away from the peak, specifically to the left leg, is thus necessary to increase the likelihood of improved selectivity. By creating histograms for each given descriptor in the domain, $c_{\ce{HER}}^{U_L}$ is then extended to be a function of $\langle \Delta G_{\ce{NNH}^*}\rangle)$. Overlaying this confidence function with the NRR activity volcano allows for visualization of this selectivity-activity trade-off (Fig. \ref{fgr:select_volc}c). Here, it is observed that indeed, moving along the stronger binding leg of the volcano shows the greatest increase in the confidence of HER suppression at the expense of activity. Considering both the expected and limiting potential volcanoes, their intersection occurs at approximately $\langle \Delta G_{\ce{NNH}^*}\rangle)=-0.5$ eV. Therefore, a material at this adsorption strength has the highest chance of suppressing HER for as little activity sacrificed as possible. Thus, a design criteria emerges, that in the pursuit of both high activity and selectivity, exclusively searching for materials that are near the NRR volcano peak is not enough, and instead we should be searching for materials on the strong binding leg as this has the highest chance for improved selectivity.

\section{Conclusions}
In this work, we present a robust methodological framework to investigate electrocatalysts towards electrochemical ammonia synthesis. We demonstrate that a Bayesian error estimation ensemble approach is capable of describing the uncertainty associated with computational parameter selection in DFT calculations for this context. Applying this framework to NRR on Fe$_1$, Ru$_1$, and Pt$_1$ we showcase its ability to generate limiting potential distributions that encompasses reported literature values. We observe that Ru$_1$ has the largest skew in its distribution implying the presence of an upper bound. The scaling relationships among the NRR intermediates \ce{N}*, \ce{NNH}*, and \ce{NH2}* are studied, and the strength of the scaling between $\Delta G_{\ce{NNH}^*}$ and $\Delta G_{\ce{NH2}^* + \ce{NH3}}$ is able to explain the upper limiting potential limit through the generation of a probabilistic activity volcano. We also apply this methodology to investigate the selectivity of these systems towards NRR, and observe the presence of a selectivity-activity trade-off. This results in a design principle that future efforts should be focused toward exploring materials on the stronger binding left leg of the volcano. Future studies could explore the single atom design space further using the procedure outlined here. Since the approach presented here requires little extra computational resources, we hope this will become a routine part of electrocatalyst design workflow. The methodological process presented here paves the way towards computational NRR works whose conclusions are robust towards the selection of parameters. We believe this will open the door towards more conservative computational studies that can provide more realistic estimates towards catalytic performance.

\begin{acknowledgement}

% Please use ``The authors thank \ldots'' rather than ``The
% authors would like to thank \ldots''.

% The author thanks Mats Dahlgren for version one of \textsf{achemso},
% and Donald Arseneau for the code taken from \textsf{cite} to move
% citations after punctuation. Many users have provided feedback on the
% class, which is reflected in all of the different demonstrations
% shown in this document.

L.K. acknowledges the support of the Natural Sciences and Engineering Research Council of Canada (NSERC). V.V. acknowledges support from the Advanced Research Projects Agency Energy (ARPA-E) under Grant \texttt{DE-AR0001211}.  L.K. thanks Dilip Krishnamurthy and Olga Vinogradova for important discussions. 

\end{acknowledgement}

\begin{suppinfo}

% This will usually read something like: ``Experimental procedures and
% characterization data for all new compounds. The class will
% automatically add a sentence pointing to the information on-line:

Computational details for generating the probabilistic activity volcanoes. $\Delta E_{\ce{NNH}^*}$ and $\Delta E_{\ce{H}^*}$ ensembles with corresponding skewness and kurtosis values. Free Energy landscapes for Fe$_1$ and Pt$_1$. Additional scaling relation plots. Probabilistic HER volcano

\end{suppinfo}

% \bibliography{achemso-demo}
\bibliography{lance_references,gpaw_ase}
% \bibliography{gpaw_ase}

\includepdf[pages=-]{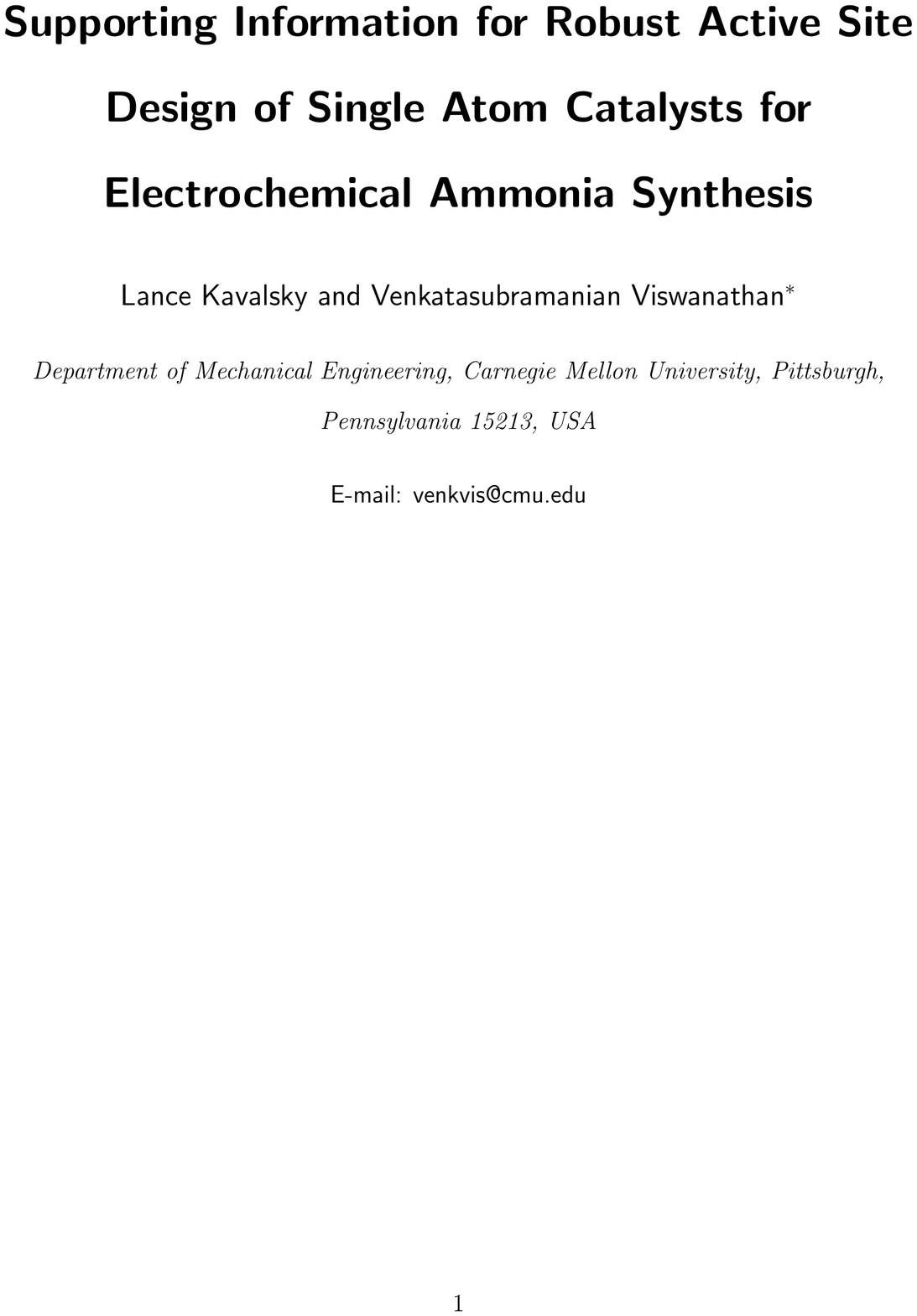}

\end{document}